\documentclass[reprint,amsmath,amssymb,aps,pre]{revtex4-1}
\usepackage[dvips]{graphicx}
\usepackage{bm}
\usepackage{mathtools}

\begin{document}
\title{
  Relaxational processes in the one-dimensional
  Ising model with long-range interactions
}
\author{Yusuke Tomita}
\affiliation{College of Engineering, Shibaura Institute of Technology, Minuma-ku, Saitama 337-8570, Japan}

\date{\today}

\begin{abstract}
  Relaxational processes in ordered phases of one-dimensional
  Ising models with long-range interactions are investigated
  by Monte Carlo simulations.
  Three types of spin model, 
  the pure ferromagnetic, the diluted ferromagnetic,
  and the spin glass models, are examined.
  The effective dimension of the one-dimensional systems are controlled
  by a parameter $\sigma$, which tunes the rate of interaction decay.
  Systematical investigations of droplet dynamics,
  from the lower to the upper critical dimension,
  are conducted by changing the value of $\sigma$.
  Comparing numerical data with the droplet theory,
  it is found that the surface dimension of droplets is distributed
  around the effective dimension.
  The distribution in the surface dimension makes the droplet dynamics
  complex and extremely enhances dynamical crossover.
\end{abstract}


\maketitle

\section{%
  \label{sec:intro}
  Introduction
}

Probing dynamical properties is an indispensable mean
to investigate magnetic and/or dielectric materials.
To extract the dynamical properties,
several methods (e.g., x-ray scattering, neutron scattering,
nuclear magnetic resonance, electron spin resonance,
muon spin resonance, etc.) are adopted in various research fields.
While a couple of one-shot experiments could extract characters of a material, 
a large number of experiments on a group of similar substances will be required
to acquire a fundamental understanding of an intriguing phenomenon.
In order to obtain an organized view on the phenomenon,
an effective framework which visualizes relations
between corresponding experiments is important.
The Cole-Cole plot~\cite{ColeCole1941} is an example of such framework.
It is a plot of the real part and the imaginary part of the permittivity
which visualizes dielectric relaxation of a substance
and makes it easier to classify dielectric materials
by comparing each Cole-Cole plot.
Another example is the Angell plot for glassforming liquids~\cite{Angell2000}.
It plots a viscosity of a substance as a function of scaled temperature,
and visualizes whether it is fragile or strong liquid.
These frameworks categorize dynamics of various substances
in a simple manner,
and we need a theory that integrates the findings of the dynamics
in many-body systems into a unified perspective.
Though dynamics at criticalities are well studied~\cite{HohenbergHalperin1977, OzekiIto2007},
dynamics in ordered phases are not clarified sufficiently.

The droplet theory is one of theories dealing with
dynamics in ordered phases of many-body systems.
The droplet theory has succeeded in explaining
dynamics of pure ferromagnetic models~\cite{HuseFisher1987, Tang1989},
ferromagnetic models with randomness~\cite{HuseFisher1987},
and the spin glass model~\cite{FisherHuse1986, FisherHuse1988}.
The droplet model assumes a shape of the free-energy landscape of droplets,
which are domains of an ordered state,
and deduces a relaxational behavior of the system.
Dynamical properties relate closely to the shape of droplets,
which depends on both the dimensionality of the system and
the type of the interactions.
To develop understanding of ordered states,
a comprehensive study on relations between dynamical properties and
the shape of droplets is required.
The one-dimensional (1D) Ising model with long-range interactions
[Eq.~(\ref{eq:hamiltonian})]
provides a suitable test ground for the comprehensive study.
Though the spin model looks simple,
it exhibits pure ferromagnetic,
ferromagnetic with randomness, and spin glass phases
by changing the distribution of interactions and
tuning the rate of interaction decay.
Tunable long-range interactions of the model enables to 
investigate the relaxational processes in a continuous manner
from the lower to the upper critical dimension.
Through numerical analyses of the 1D Ising model,
we examine the droplet theory as a suitable framework
to describe the mechanism of various dynamics in many-body spin systems.

This paper is organized as follows:
In Sec.~\ref{sec:review}, we briefly review preceding studies
on the 1D Ising model with long-range interactions
and the droplet theory.
Details on numerical calculations are given in Sec.~\ref{sec:method}.
Section~\ref{sec:results} presents results obtained
by Monte Carlo simulations of the 1D Ising model.
Section~\ref{sec:discussion} is devoted to the discussion.
The summary of this paper is presented in Sec.~\ref{sec:summary}.
A detail explanation of $O(N)$ Monte Carlo method~\cite{FukuiTodo2009,Tomita2009a,Tomita2009b},
which is a key algorithm to achieve numerical calculations
of systems with long-range interactions with reasonable computational cost,
is given in Appendix~\ref{sec:appendix}.

\section{%
  \label{sec:review}
  Reviews of the 1D Ising model with long-range interactions and
  the droplet theory
}

This section presents brief reviews on the 1D Ising model with
long-range interactions and the droplet theory.

The Hamiltonian of the 1D Ising model is given by
\begin{equation}
  \label{eq:hamiltonian}
  {\cal H} = -\sum_{i < j}\frac{J_{ij}}{r_{ij}^{\sigma}}S_iS_j.
\end{equation}
Here, $S_i(\in \{1, -1\})$ represents the Ising spin variable at site $i$,
$J_{ij}$ is the exchange interaction between $i$ and $j$,
$r_{ij}$ is the distance between $i$ and $j$,
and $\sigma$ is the tuning parameter of long-range interaction.
Despite its dimensionality and simple appearance,
the model possesses various features which are controlled by
the model parameters:
the complexity of the interactions, $J_{ij}$, and
the tuning parameter of the interaction decay, $\sigma$.

The section consists of three subsections.
In the first subsection, we review preceding studies
on pure ferromagnetic models.
The second subsection gives reviews on diluted ferromagnetic models.
Spin glass models are reviewed in the third subsection.

\subsection{%
  \label{sec:review_pure}
  Pure ferromagnetic model
}

The pure ferromagnetic model is characterized by
a uniform ferromagnetic interaction:
\begin{equation}
  J_{ij} = J(> 0) \quad \mbox{(for any $i$ and $j$).}
\end{equation}
The interaction decay with increasing distance between spin pairs
is tuned by the parameter $\sigma$:
The long-range interaction is irrelevant when $\sigma > 2$,
and the system does not exhibit the ferromagnetic order at
any finite temperature.
The critical phenomena of the system belong to 
the universality class of the mean-field model, when $\sigma \le 3/2$.
At $\sigma = 2$, the Kosterlitz-Thouless (KT) transition~\cite{Thouless1969, AndersonYuval1971, Kosterlitz1976}
appears at a finite temperature.
In the range $3/2 < \sigma < 2$,
the universality class of the ferromagnetic phase transition
depends on the value of $\sigma$.
Therefore, by varying $\sigma$ from 2 to 3/2,
we can continuously survey the Ising model
from the lower critical [$d(\sigma=2) = 1$]
to the upper critical [$d(\sigma=3/2) = 4$] dimension.
It should be noticed that the 1D Ising model on the trace of
the effective dimension [$d(\sigma)$]
does not correspond to the nearest-neighbor model.
For example, the 2D nearest-neighbor model exhibits
a logarithmic divergence of the specific heat
while the critical exponent of the specific heat
of the 1D model is positive in the range of $3/2 < \sigma < 2$.

While we have little analytical results on dynamics of the Ising model,
rigorous results of dynamics in the paramagnetic phase are
given by Glauber~\cite{Glauber1963}.
Using his results,
the time-delayed correlation function $C(k; t)$ is given by
\begin{align}
  \label{eq:cf_ferro}
  C(k; t) &= \langle \tilde{S}_{-k}(0)\tilde{S}_k(t)\rangle \\
  \label{eq:cf_para}
  &= \xi(k)\exp[-t/\tau(k)],
\end{align}
where $\tilde{S}_k(t)$ is the Fourier transform of the Ising spin at time $t$,
$\xi(k)$ is the correlation length for the wave number $k$,
$\tau(k)$ is the lifetime for the wave number $k$,
and $T$ is the temperature
(the Boltzmann constant $k_{\rm B}$ is set to unity).
The angle brackets $\langle \cdots \rangle$ denote a thermal average.
The Fourier transform of the Ising spin $\tilde{S}_k(t)$,
the correlation length $\xi(k)$, and the lifetime $\tau(k)$
are, respectively, given by
\begin{align}
  \tilde{S}_k(t) &= \frac{1}{L}\sum_{r}S_r(t)e^{ikr},\\
  \xi(k) &= [\cosh (2J/T)(1 - \gamma\cos k)]^{-1},\\
  \tau(k) &= [a(1 - \gamma\cos k)]^{-1},
\end{align}
where $L$ is the number of spins,
$\gamma = \tanh 2J/T$, and $a$ is a nonuniversal constant.
The result shows that the correlation function in the paramagnetic phase
decays exponentially with time, and the lifetime is proportional
to the correlation length.

The correlation function of the Ising model in the mean-field region
is given by several authors~\cite{SuzukiKubo1968, KawasakiYamada1968, Abe1968}.
The time-delayed correlation function $C(t)$ in the ordered phase is
\begin{align}
  C(t) &= \langle S(0)S(t)\rangle \nonumber \\
  \label{eq:cf_mf}
  &\sim \frac{C_0^2}{(C_0/C_\infty)^2 - [(C_0/C_\infty)^2-1]e^{-t/\tau}},
\end{align}
where $C_0$ and $C_\infty$ are nonuniversal constants.
The result shows that
the correlation function in the ordered phase shows
an exponential convergence as well as in the paramagnetic phase.

In the intermediate dimension ($1 < d < 4$),
there is no rigorous result of dynamics in the ordered phase,
but deduced forms of the autocorrelation function
by the droplet theory are available.
When the spatial dimensionality $d$ is sufficiently low ($d < 3$)
in the nearest-neighbor interaction model,
Huse and Fisher have shown that the autocorrelation function $C_i(t)$
at site $i$
shows the Kohlrausch-Williams-Watts stretched exponential decay
as~\cite{HuseFisher1987},
\begin{align}
  \label{eq:KWW}
  C_i(t) &= \langle S_i(0)S_i(t)\rangle - \langle S_i\rangle^2\nonumber \\
  &\sim \exp[-(t/\tau)^{(d-1)/2}].
\end{align}
This stretched exponential decay comes from the emergence of large-scale
droplets.
Since a lifetime of large-scale droplet is long,
the excited droplets dominate the dynamics of the system.
On the other hand, such large-scale droplets do not emerge
at a high dimensionality.
Larger droplets are much more affected by thermal fluctuations
since the surface area of droplets increases proportionally with
the linear size $l$ to the power of $d-1$, $l^{d-1}$.
For $d > 3$, fluctuations of average size droplets dominate
the correlations, and the system shows a simple exponential decay,
$C_i(t) \sim \exp[-(t/\tau)]$.
The prediction of the droplet theory indicates
that the 1D Ising model will show a simple exponential decay
in high enough dimension.
In other words, there is a critical value of
$\sigma_c$ [$d(\sigma_c) = 3$]
where the form of the autocorrelation function changes.

\subsection{%
  \label{sec:review_dilute}
  Diluted ferromagnetic model
}

The diluted ferromagnetic model possesses randomness without frustration,
and it serves a suitable test ground to study effects of randomness.
There are two types of dilution, site- and bond-dilution.
In the present paper, we deal with a bond-dilution model
whose Hamiltonian is given by Eq.~(\ref{eq:hamiltonian})
with randomly diluted interactions,
\begin{equation}
  J_{ij} = \left\{
  \begin{array}{ll}
    J(> 0)     & \mbox{(with probability 1/2),}\\
    J'(= 0) & \mbox{(with probability 1/2).}
  \end{array}
  \right.
\end{equation}
The value of the diluted interaction is
chosen so as to maximize effects of randomness.
The disconnections of the interaction ($J' = 0$) bring a lowering
of a transition temperature,
and the ordered state in a low temperature
is strongly affected by the geometry of the interaction network.
The geometrical effect is stronger at larger $\sigma$
since the phase transition temperature decreases as increases $\sigma$.
The details of the effect is discussed in Sec.~\ref{sec:discussion}.

The phase diagram of the diluted model ($J' \ge 0$)
is almost the same as the pure model.
However, according to the Harris criterion~\cite{Harris1974},
the universality class of the diluted model in the range of $3/2 < \sigma < 2$,
where the critical exponent of the specific heat of the pure model $\alpha$
is positive, is altered by the dilution.
On the other hand, the critical exponent $\alpha$ is zero
when $\sigma = 2$ and $\sigma \le 3/2$,
and the dilution will be irrelevant to the universality class.
For systems that exhibit the KT transition,
it will not be a trivial question whether the dilution
alters the universality class or not,
but it seems irrelevant regarding
numerical studies of the two-dimensional
diluted XY model~\cite{Berche2003, SurunganOkabe2005}.

By assuming the dilution being not so strong,
Huse and Fisher derived that the autocorrelation function shows
a power-law decay,
\begin{equation}
  \label{eq:cf_dil_droplet}
  \overline{C_i}(t) \sim t^{-x(T)},
\end{equation}
where the overbar denotes the sample average of the interaction realization.
The exponent $x(T)$ will depend on the temperature
and nonuniversal details of the system~\cite{HuseFisher1987}.
The arrangement of diluted bonds could strongly affect
the relaxation at low temperatures ($T \ll T_c$) since the exponent $x(T)$
depends on the rate of the creation and the annihilation of large droplets.
However, thermal fluctuations blur the details of the arrangement
of diluted bonds,
and $x(T)$ presumably approaches a universal value
near the transition temperature. 


\subsection{%
  \label{sec:review_sg}
  Spin glass model
}

Interactions of the spin glass model consist of ferromagnetic and
antiferromagnetic bonds, and the random arrangement of bonds
brings about frustration in the system.
In the present paper, we deal with a random bond model
whose Hamiltonian is given by Eq.~(\ref{eq:hamiltonian})
with randomly mixed interactions,
\begin{equation}
  J_{ij} = \left\{
  \begin{array}{ll}
    J & \mbox{(with probability 1/2),}\\
    -J & \mbox{(with probability 1/2).}
  \end{array}
  \right.
\end{equation}
Though the bimodal distribution in the exchange interaction
does not bring about frustration in the 1D model
when nearest-neighbor interactions are dominance,
frustration emerges when interactions are sufficiently long-ranged.
Kotliar, Anderson, and Stein showed that
the spin glass phase appears when $\sigma < 1$:
the universality class of the model belongs to
that of the mean-field for $\sigma < 2/3$,
and that depends on $\sigma$ for $2/3 < \sigma < 1$~\cite{Kotliar1983}.

As in the diluted model,
the relaxation of the system is altered further by frustrated interactions.
Fisher and Huse derived the logarithmic decay of autocorrelation function,
\begin{equation}
  \label{eq:cf_sg_droplet}
  \overline{C}(t) \sim [\ln(t/\tau_0)]^{-\phi},
\end{equation}
where $\tau_0$ and $\phi$ are, respectively,
a microscopic time scale and a nontrivial exponent~\cite{FisherHuse1986, FisherHuse1988}.
The logarithmic decay originates from the distributions of
the droplet free energy $F_L$ and the droplet barrier for annihilation
of droplets $B_L$.
Both of the two distributions have broad distribution
and different size dependencies,
$F_L \sim L^{\theta}$ and $B_L \sim L^{\psi}$.
The exponent $\phi$ in Eq.~(\ref{eq:cf_sg_droplet}) is the ratio of
the exponents, $\phi = \theta/\psi$.

%
%

In the mean-field regime,
the time-delayed correlation function at site $i$ in the spin glass phase,
$\overline{C}_i(t)$,
is given by~\cite{KirkpatrickSherrington1978}
\begin{equation}
  \overline{C}_i(t) = \frac{1-C_0}{(1+at)^{1/2}} + C_0,
\end{equation}
where $a$ and $C_0$ are constants.
Unlike the case of the ferromagnetic model,
the correlation function shows the power-law decay
with an exponent 1/2.

\section{%
  \label{sec:method}
  Method
}

While the 1D Ising model with long-range interactions
is a suitable spin model for analytical studies,
the large computational cost of scanning all the interactions
proportional to the square of the system size, $O(L^2)$,
hampers numerical studies.
To overcome the numerical difficulty in systems with
long-range interactions,
Fukui and Todo proposed $O(N)$ cluster Monte Carlo (MC) method~\cite{FukuiTodo2009}.
The $O(N)$ cluster MC method was successfully applied to
study the single-spin-flip dynamics
in 1D Ising models with power law decaying interactions
and nontrivially frustrated systems~\cite{Tomita2009a}, and
2D Heisenberg dipolar lattices~\cite{Tomita2009b}.
The details of the algorithm are given in Appendix.

To eliminate the edges of the system,
the periodic boundary condition is imposed
to all simulations.
For the pure and the diluted ferromagnetic models,
all of the contributions from supercells are summed up.
When $\sigma > 1$, the summation is easily executed as
\begin{align}
  \frac{1}{\tilde{r}_{ij}^\sigma} &=
  \sum_{n=-\infty}^{\infty}\frac{1}{(r_{ij} + nL)^\sigma} \nonumber\\
  &= \frac{1}{L^{\sigma}}
      [\zeta(\sigma, r_{ij}/L) + \zeta(\sigma, (L - r_{ij})/L)],
\end{align}
where $\tilde{r}_{ij}$ and $\zeta(\sigma, r)$
are, respectively, the effective distance between site $i$ and $j$
and the Hurwitz zeta function,
\begin{equation}
  \zeta(\sigma, r) = \sum_{n=0}^{\infty}\frac{1}{(n + r)^\sigma}.
\end{equation}
However, the summation does not converge when $\sigma \le 1$.
In order to implement the periodic boundary condition
when $\sigma \le 1$,
the chord distance~\cite{KatzYoung2003} is implemented
in the spin glass model:
We place a spin $S_i$ equidistantly on a ring of length $L$,
and the distance $r_{ij}$ is described by
\begin{equation}
  \label{eq:chord}
  r_{ij} = \frac{L}{\pi}\sin\left(
  \frac{\pi |i-j|}{L}
  \right).
\end{equation}
The slow decay of the interaction causes a large transition temperature.
Therefore, we rescale the interaction $J$ to $c(\sigma, L)J$ as
\begin{equation}
  c(\sigma, L) = \sqrt{\frac{L}{2J\sum_{i < j}r_{ij}^{-2\sigma}}}.
\end{equation}
This rescaling adjusts the transition temperature at $\sigma = 0$ to unity
and makes the transition temperature to be moderate for $\sigma > 0$.
In the MC simulation, the interactions considered above are employed.

To investigate the dynamical properties, we calculate
time-delayed correlation functions.
In the pure and the diluted ferromagnetic models,
the correlation function is calculated by Eq.~(\ref{eq:cf_ferro}).
The time-delayed correlation function in the spin glass model is
defined as
\begin{equation}
  \label{eq:cf_sg}
  \overline{C}(k; t) = 
    \overline{\langle \tilde{q}_{-k}(0)\tilde{q}_k(t)\rangle},
\end{equation}
where $\tilde{q}_k(t)$ is the Fourier transform of
the Edwards-Anderson order parameter~\cite{EdwardsAnderson1975} at time $t$,
which is given by
\begin{equation}
  \label{eq:edwards_anderson}
  \tilde{q}_k(t) = \frac{1}{L}\sum_{r}S^{(1)}_r(t)S^{(2)}_r(t)e^{ikr}.
\end{equation}
Here, the upper suffixes $(1)$ and $(2)$ denote
the replica indexes.
The time-delayed correlation function [Eq.~(\ref{eq:cf_sg})] is
different from the autocorrelation function considered
in the droplet theory~\cite{FisherHuse1986, HuseFisher1987, FisherHuse1988}.
Whereas the droplet theory deals with the time-delayed correlation
of a local spin at site $i$,
Eq.~(\ref{eq:cf_sg}) deals with that of the macroscopic
order parameter.

\section{%
  \label{sec:results}
  Results
}

In this section, results obtained by Monte Carlo simulations
on three different models,
pure ferromagnetic, dilute ferromagnetic, and
spin glass, are shown.
In what follows, we use $J$ as a unit of temperature.
Error bars of obtained data are omitted
since large error bars impair the visibility of figures.
Large statistical errors of the data
mainly come from small values of observables.
Though statistical errors at each point are large,
trends of decay in autocorrelations are clearly observed
as we see below.

\subsection{%
  \label{sec:result_pure}
  Pure ferromagnetic model
}

This subsection gives results obtained by Monte Carlo simulations
of the 1D pure ferromagnetic model.
Monte Carlo simulations are executed for several values of $\sigma$,
a parameter of long-range interactions.
The system size is set as $L = 2^{20} (\simeq 1.0\times 10^6)$.
The system is equilibrated by
combination use of annealing and a cluster flip update~\cite{FukuiTodo2009}.
10$^7$ Monte Carlo steps are executed
for measurement of autocorrelation functions,
and 10 independent samples are simulated for obtaining good statistics.

\begin{figure}[h]
  \includegraphics[width=.4\textwidth]{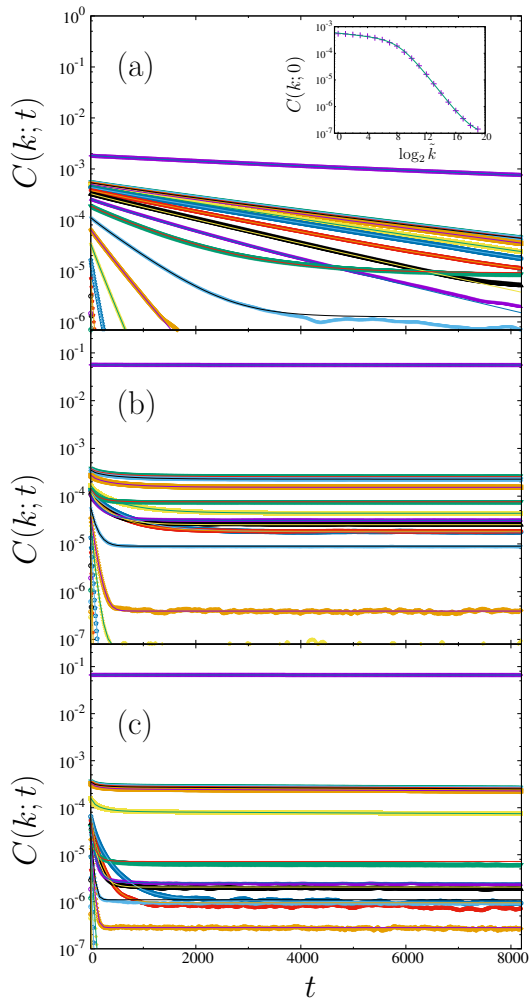}
  \caption{\label{figs:ferro}
    (Color online) Autocorrelation functions of the pure ferromagnetic model
    at (a) $\sigma = 2$, (b) $\sigma = 1.8$,
    and (c) $\sigma = 1.6$.
    Topmost lines in each figure are
    autocorrelation functions of $k = 0$.
    Following to the function of $k = 0$,
    autocorrelation functions of $\log_2\tilde{k} = 0, 1, 2, \ldots$
    align from top to bottom in almost ascending order.
    Inset shows $k$-dependence of the autocorrelation functions
    at $\sigma = 2$ and $t = 0$.
  }
\end{figure}

As reviewed in Section~\ref{sec:review_pure}, the pure ferromagnetic model
shows the KT transition at a finite temperature when $\sigma = 2$.
Autocorrelation functions would exhibit simple exponential decays
even below the transition temperature
since there is no true long-range order.
Unlike those in normal paramagnetic phase,
the autocorrelation function in the KT phase at time $t$, $C(k; t)$,
depends on a power of a wave number $k$,
\begin{equation}
  C(k; t) \propto k^{-\eta}\exp[-t/\tau(k)],
  \label{eq:relax_kt}
\end{equation}
where $\eta$ is the critical exponent of the correlation function
and $\tau(k)$ is the lifetime for $k$.
The autocorrelation functions in the KT phase ($T=1.78$) obtained
by Monte Carlo simulation are plotted in Figs.~\ref{figs:ferro}(a).
Wave numbers in Figs.~\ref{figs:ferro}(a) are
zero and $\tilde{k}=2^n (n = 0, 1, 2,\ldots, 19)$,
where $\tilde{k}$ denotes scaled wave number, $\tilde{k} = Lk/2\pi$.
Curves of the autocorrelation functions at $t = 0$ (MCS$=0$)
ought to be equally-spaced in the logarithmic
scale if their relaxations are described by Eq.~(\ref{eq:relax_kt}).
But Figs.~\ref{figs:ferro}(a) shows distances between neighboring curves
become wider as $\tilde{k}(\ge 1)$ increases.
To examine the $k$-dependence of autocorrelation functions at $t = 0$,
I assumed a $k$-dependence,
\begin{equation}
  C(\tilde{k}; 0) = C_1\frac{k^{-\eta}}{1 + (k/\kappa)^\phi} + C_2,
  \label{eq:mod_oz}
\end{equation}
rather than Eq.~(\ref{eq:relax_kt}).
Constants $C_1$, $C_2$, and $\kappa$ are nonuniversal constants.
It is known that $\eta = 0$ and $\phi = 2$
when the system is in the mean field region~\cite{GebhardtKrey}.
The formula, Eq.~(\ref{eq:mod_oz}), conforms to Eq.~(\ref{eq:relax_kt})
at $t = 0$ when $\phi = 0$.
The result is shown in inset of Figs.~\ref{figs:ferro}(a).
Estimated values are $\eta = 0.10$, $\phi = 1.1$,
and $\kappa = 2.0\times 10^{-3}$.
The result indicates that the system seems to be in the KT phase
in the range of $k < \kappa$ whereas the system seems to be in
an ordered state in the range of $k > \kappa$.
The emergence of the ordered state can be explained by
the characteristic length of ferromagnetic clusters of Ising spin.
At a low enough temperature, larger ferromagnetic clusters are more
stable than smaller ones.
Because of the discreteness of the Ising spin,
there is little cluster whose size is smaller than $1/\kappa$
at such a low temperature.
Therefore the system seems ordered in the range of $k > \kappa$.

As mentioned in Sec.~\ref{sec:review_pure},
the relaxation of autocorrelation function depends on
the dimensionality of the droplet,
which is directly affected by the parameter $\sigma$.
According to the results by Tang, Nakanishi, and Langer~\cite{Tang1989},
there will be a critical parameter $\sigma_c$:
A stretched exponential decay will be observed when $\sigma_c < \sigma < 2$,
while a relaxation will be simple exponential decay when $\sigma < \sigma_c$.
Figures~\ref{figs:ferro}(b) and (c) show exponential decays
observed in an ordered phase $T = 2.48$ at $\sigma = 1.8$ and
$T = 3.50$ at $\sigma = 1.6$, respectively.
In order to examine forms of relaxations of order parameter,
multiparameter fittings are performed while assuming a fitting form,
\begin{equation}
  C(k; t) = C_0\exp[-(t/\tau(k))^{\beta(k)}] + C_{\infty},
  \label{eq:str_exp1}
\end{equation}
where $\beta(k)$ is a stretching exponent for $k$
and $C_0$ and $C_{\infty}$ are nonuniversal constants.
The fitting form Eq.~(\ref{eq:str_exp1}) well fits data
when $\sigma \ge 1.7$, but the form is not good enough to fit data
when $\sigma \le 1.6$.
By adding another exponential term to the form,
fitting results are fairly improved.
The fitting form applied to systems with $\sigma \le 1.6$ is
\begin{align}
  C(k; t) &= C_0\exp[-(t/\tau(k))^{\beta(k)}]\nonumber\\
  &\quad + C_2\exp[-(t/\tau_2(k))^{\beta_2(k)}] + C_{\infty},
  \label{eq:str_exp2}
\end{align}
where $\beta(k)$ and $\beta_2(k)$ are stretching exponents for $k$,
$\tau(k)$ and $\tau_2(k)(>\tau(k))$ are lifetimes for $k$,
and $C_0$, $C_2$, and $C_{\infty}$ are nonuniversal constants.
The necessity of two exponential terms means that
there is another non-negligible 
mode in relaxational processes when $\sigma \le 1.6$.
The origin of the two non-negligible modes is discussed in
Sec.~\ref{sec:discussion}.

\begin{figure}[h]
  \includegraphics[width=.4\textwidth]{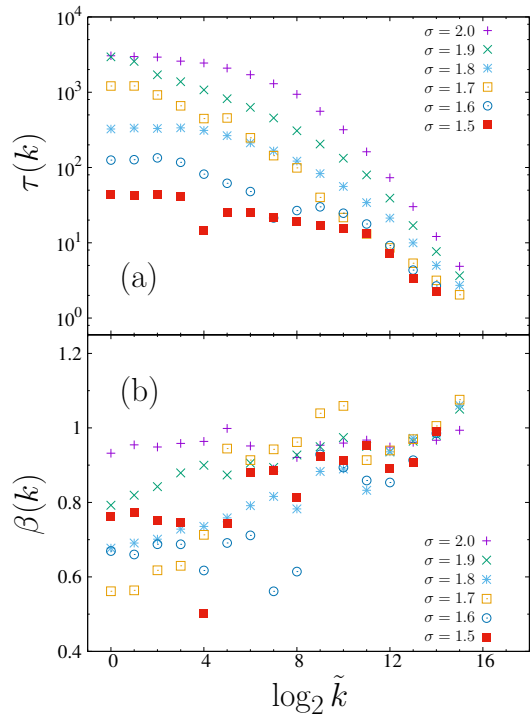}
  \caption{\label{figs:ferro_bt}
    (Color online) (a) Lifetimes for several $\sigma$'s
    as functions of $\tilde{k}$.
    Lifetimes tend to decrease as $\tilde{k}$ increases;
    lifetimes of small droplets are shorter than those of larger ones.
    (b) Stretching exponents for several $\sigma$'s
    as functions of $\tilde{k}$.
    The exponents are nearly continuous functions of $\tilde{k}$
    when $\sigma \ge 1.8$, whereas they have a discontinuous jump
    at $\tilde{k} \sim 5$ when $\sigma \le 1.7$.
  }
\end{figure}

The stretching exponents and lifetimes are plotted in Figs.~\ref{figs:ferro_bt}.
The exponents $\beta(k)$ at $\sigma = 2.0$ are almost independent of $k$.
The value of $\beta$ is unity when the long-range order is absent.
Therefore values of $\beta(k)$ are expected to be unity since the KT phase
does not have the true long-range order.
The obtained values of $\beta(k)$ are, however, slightly smaller
than unity. This would be caused by the finiteness of the system size.
The correlation length is larger than the system size, so that
the system seems as if it has true long-range order.

The values of $\beta(k)$ at $\sigma = 1.9$ and $\sigma = 1.8$
increase and asymptotically approach unity as $k$ increases.
According to the droplet theory,
$\beta$ is an increasing function of the dimensionality,
$\beta = (d-1)/2$.
That is, the value of $\beta$ ought to be small at $\sigma \lesssim 2$,
but numerical data are inconsistent with the droplet theory.
This inconsistency is discussed in Sec.~\ref{sec:discussion}.

The dependence of $\beta$ on $k$ is altered between $\sigma = 1.8$ and 1.7.
While $\beta$ changes continuously with $k$ when $\sigma \ge 1.8$,
$\beta$ shows a discontinuous change at a certain point when $\sigma \le 1.7$.
This intrinsic change in the $k$-dependence of $\beta$
corresponds to an outcome of the droplet theory:
The density of relaxational modes switches from continuous to
discrete at $d=3$ as increasing $d$.
The droplet theory also declare that the relaxation is a simple
exponential when $d > 3$.
The discrepancy between the simple exponential and
stretched exponential decays in the systems at $\sigma \le 1.7$
is discussed in Sec.~\ref{sec:discussion}.

\subsection{%
  \label{sec:result_dilute}
  Diluted ferromagnetic model
}

This subsection gives results of the 1D diluted ferromagnetic model.
The system size $L$ and the parameter $\sigma$'s are
the same as in the pure ferromagnetic model.
The equilibration method is also the same;
the combination use of annealing and a cluster flip update is employed.
The number of Monte Carlo steps is reduced to 10$^6$ steps per sample,
while the number of random samples is increased to
100 for $\sigma = 2.0, 1.9, 1.8, 1.7$ and 1.4 and
200 for $\sigma = 1.6$ and 1.5, respectively.

As mentioned in Sec.~\ref{sec:review_dilute},
we expect that the topology of the phase diagram is the same as
that of the pure ferromagnetic model.
Though the ordered phase is simple ferromagnetic,
the dilution could change dynamical properties of the model.
In fact, Huse and Fisher showed that
excited large ferromagnetic droplets emerged by quenched bond disorder
bring about a power-law decay~\cite{HuseFisher1987}.
The relevance of the bond dilution can be estimated
by the Harris criterion~\cite{Harris1974}.
Since, based on the Harris criterion,
the critical exponent of the specific heat $\alpha$ of
the pure ferromagnetic model is positive in $3/2 < \sigma < 2$~\cite{Fisher1972, Suzuki1973, Tomita2009a},
autocorrelation functions are expected
to show the power-law decay in the range of $\sigma$.
On the other hand, we expect the exponential decay
in the mean-field region ($\sigma \le 3/2$) and
the KT phase ($\sigma = 2$)
because the critical exponent $\alpha$ is zero in the region of $\sigma$.

\begin{figure}[h]
  \includegraphics[width=.4\textwidth]{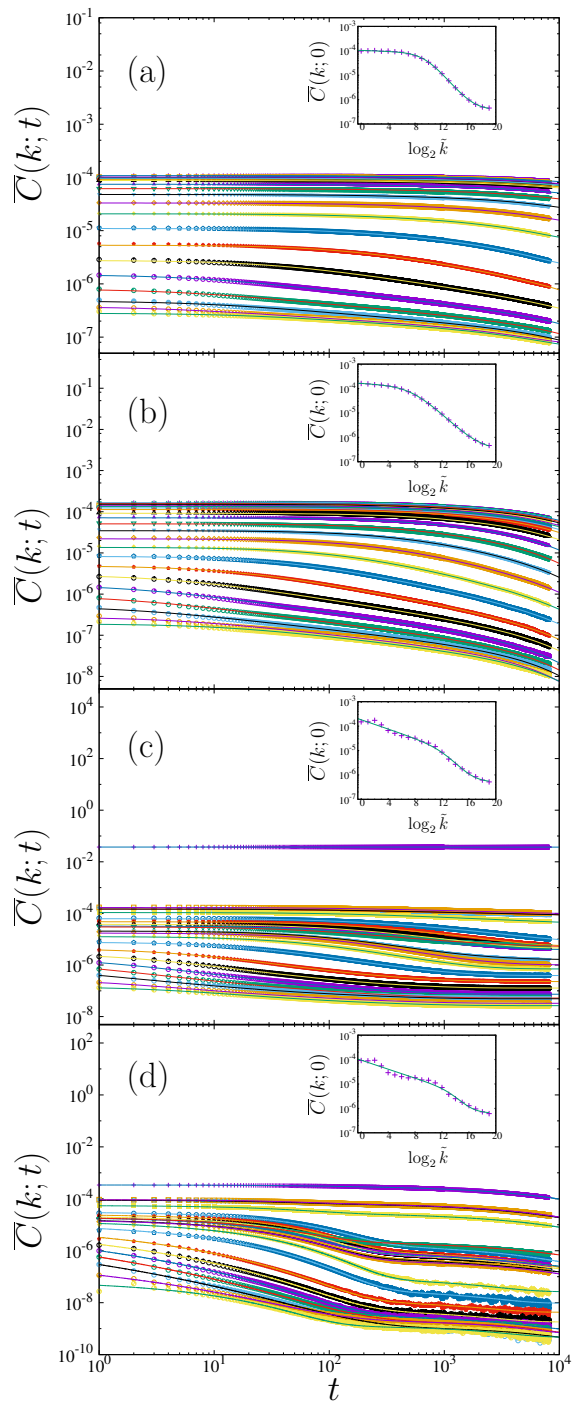}
  \caption{\label{figs:dilute}
    (Color online) Autocorrelation functions of the diluted ferromagnetic model at
    (a) $\sigma = 2$,
    (b) $\sigma = 1.8$,
    (c) $\sigma = 1.6$, and
    (d) $\sigma = 1.4$.
    Insets show $k$-dependence of the autocorrelation functions at $t = 0$.
  }
\end{figure}

Autocorrelation functions at $\sigma = 2.0$, 1,8, 1.6, and 1.4 are plotted in
Figs.~\ref{figs:dilute}. Using a fitting function,
\begin{equation}
  C(k; t) = C_0\frac{\exp(-t/\tau_2(k))}{(1 + t/\tau_1(k))^{x(k)}} + C_\infty,
  \label{eq:power}
\end{equation}
I estimate characteristic times, $\tau_1(k)$ and $\tau_2(k)$, and
the exponent of the power-law function, $x(k)$.
The parameter $\tau_1(k)$ provides an indication of the waiting time
for starting the power-law decay.
As Figs.~\ref{figs:dilute} show that autocorrelation functions of small $k$
decrease little at the beginning, and they start exhibiting power-law decay
at $t \sim O(\tau_1)$.
The introduction of the waiting time $\tau_1(k)$ in Eq.~(\ref{eq:power})
overcomes the difficulty to fit the data which show power-law decay
after a certain waiting time.
The parameter $\tau_2(k)$ is the thermal relaxation time of the autocorrelation
function;
the function nearly reaches thermally equilibrium
value $C_{\infty}$ at $t \sim O(\tau_2)$.
Autocorrelation functions at $\sigma \ge 3/2$ are well fitted by
Eq.~(\ref{eq:power}).

The critical exponent $\alpha$ is zero at $\sigma = 2$~\cite{Fisher1972, Suzuki1973},
so that the random dilution is irrelevant and does not change
its universality class, Kosterlitz-Thouless phase,
according to the Harris criterion~\cite{Harris1974}.
But the autocorrelation functions are apparently different from
the pure system at $\sigma = 2$.
And the data are well fitted by the power-law fitting form, Eq.~(\ref{eq:power}).
It seems inconsistent with the analytical results
but it is consistent if we consider that the long-range order
exists in the pure system at $\sigma = 2$ due to the finite-size effect.
Therefore, 
the finiteness of the system brings about pseudo long-range order
at the marginal point ($\sigma = 2$),
and the dilution gives rise to the power-law decay of
pseudo long-range order.

\begin{figure}[h]
  \includegraphics[width=.4\textwidth]{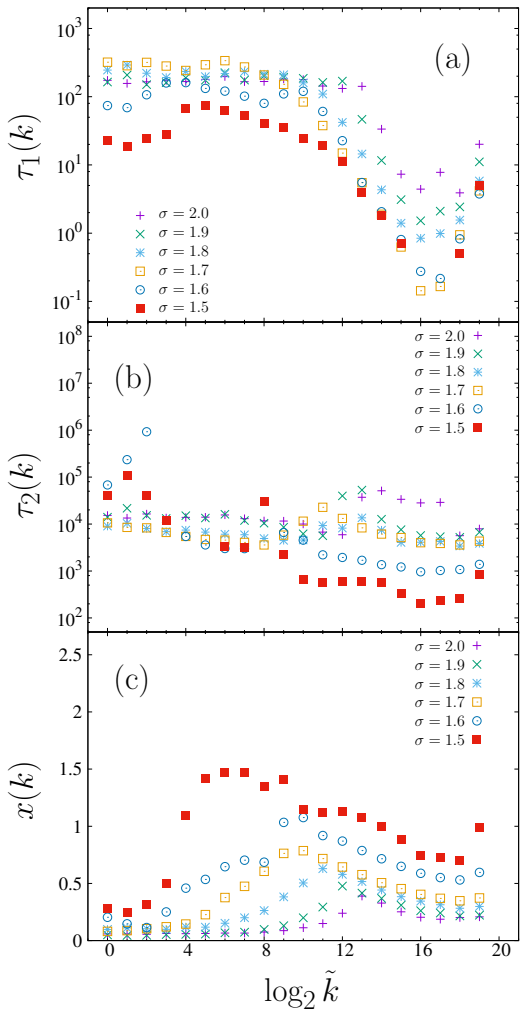}
  \caption{\label{figs:dilute_ttx}
    (Color online) (a) Waiting times $\tau_1$ for several $\sigma$'s as
    functions of $\tilde{k}$.
    At the points $\tilde{k} = 16$ and 17 data of $\tau_1$ of $\sigma = 1.5$
    are not plotted since they are almost zero.
    (b) Lifetimes $\tau_2$ for several $\sigma$'s as functions of $\tilde{k}$.
    (c) Exponents of the power-law decay for several $\sigma$'s as
    functions of $\tilde{k}$.
  }
\end{figure}

The parameters $\tau_1(k)$, $\tau_2(k)$, and the power-law exponent $x(k)$
are plotted in Figs.~\ref{figs:dilute_ttx}.
There are autocorrelation functions that hardly decay
within the prepared time window, and
estimated values of $\tau_2(k)$ of such functions
are larger than $10^4$.
The estimated values of $\tau_2$ larger than $10^4$ are not precise,
and it means that the exponential term in Eq.~(\ref{eq:power})
have little contribution to the fitting analysis.

Both $\tau_1(k)$ and $x(k)$ are small when $\tilde{k}$ is large
($\tilde{k} \ge 16$).
The reason of small $\tau_1(k)$ is that small droplets in metastable state
do not need long time to surmount free energy barrier
and immediately start power-law relaxation to the stable states.
Such small droplets frequently come out,
and the autocorrelation functions of small $\tilde{k}$ count up
multiple power-law decays with various onset time.
This overlap of droplet relaxations causes small power-law exponent $x(k)$
at large $\tilde{k}$.

The parameter $\tau_1(k)$ is large when $\tilde{k}$ is small
($\tilde{k} \le 4$).
This result is 
reasonable because the lifetime of metastable states of large droplets
is long.
Excited large droplets, therefore, do not relax readily,
and we observe autocorrelations of small $k$ are almost flat in the
range of $t < O(\tau_1)$.
The multiple parameter fitting with Eq.~(\ref{eq:power}) erroneously
infers that the exponent $x(k)$ is small when $\tilde{k}$ is small.
The reason of the improper inference is that
the autocorrelation functions hardly exhibit power-law decay
within the time window, and the exponent is estimated to be smaller
than real value.
In fact, the autocorrelation functions of small $\tilde{k}$
exhibit power-law decay with $x(k) \sim 1$ when the system size is small enough
\footnote{%
to be published in Ferroelectrics
}.

A marked feature of the autocorrelation functions of the diluted system
is that they are placed with roughly equal intervals.
Since scaled wave numbers $\tilde{k}$ are chosen from the power-of-two,
the functions are approximately described by
\begin{equation}
  C(\tilde{k}; t) \sim \tilde{k}^{-y}C(\tilde{k}=1; t),
\end{equation}
where $y$ is a constant.
The fact that the autocorrelation function is roughly proportional
to the power of $\tilde{k}$ means excited droplets have a fractal-like structure.
The autocorrelation functions at $t=0$ are plotted
the insets of Figs.~\ref{figs:dilute} as functions of $\tilde{k}$.
It can be seem from the insets that the profiles are quite
similar to that of the pure system at $\sigma = 2$ which
is located in KT phase.
However, the origin of the KT phase like feature in diluted systems
is different from that of the pure system.
Details of the origin will be discussed in Sec.~\ref{sec:discussion}.

The autocorrelation functions at $\sigma = 1.4$ (Figs.~\ref{figs:dilute}(d))
do not exhibit
power-law decay but the functions are well described by
the sum of two stretched exponential functions [Eq.~(\ref{eq:str_exp2})].
The change from the power-law to the stretched exponential decay
is expected
since randomly placed diluted interactions are averaged out
in the mean-field limit.
Contrary to the relaxation process, the autocorrelation functions
still have a fractal-like structure.
It means that the shape of excited droplets still keeps
a fractal-like structure though their dynamics are altered.

\subsection{%
  \label{sec:result_sg}
  Spin glass model
}

The results of the spin glass model is presented in this subsection.
The system size and Monte Carlo steps are significantly reduced
since the decay of the long-range interactions is slow
(the range of the interaction parameter is $\sigma \le 1$)
comparing to the pure and diluted ferromagnetic models.
The system size is set as $L = 2^{13}(= 8192)$.
During the equilibration, spins are updated by the single-spin-flip algorithm
because the simple cluster flip update does not work
in frustrated spin systems.
To accelerate relaxation
the replica exchange Monte Carlo method is used~\cite{HukushimaNemoto1996}.
For the measurement of autocorrelation functions,
10$^6$ Monte Carlo steps are executed.
The number of random samples is 100.

\begin{figure}[h]
  \includegraphics[width=.4\textwidth]{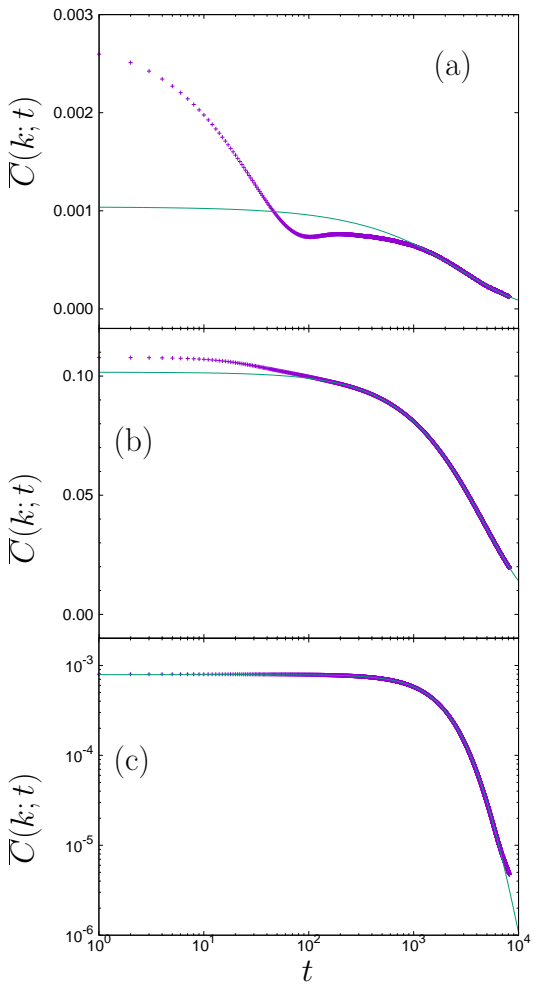}
  \caption{\label{figs:sg}
    (Color online) Autocorrelation functions of the spin glass model at
    (a) $\sigma = 0.9$,
    (b) $\sigma = 0.7$, and
    (c) $\sigma = 0.6$.
    Stretching exponential curves in (a) and (b) are fitting results of
    the $\alpha$-relaxation regime.
    Curve in (c) is obtained by using the Havriliak-Negami type
    fitting function.
  }
\end{figure}

At the lower critical value of $\sigma$(=1), Moore showed that
there is no spin glass phase at a finite temperature~\cite{Moore2010}.
However we observe a pseudo spin glass transition
at a finite temperature which seems to be independent from the system size
since the correlation length diverges extremely rapidly as lowering
the temperature.
Figures~\ref{figs:sg} show the autocorrelation functions at $\sigma = 0.9$,
0.7, and 0.6.
Autocorrelation functions of $k > 0$ are omitted
since they are fairly small comparing to that of $k = 0$.
The autocorrelation function at $\sigma = 0.9$ shows typical spin-glass behaviors~\cite{BinderKob}: 
It shows a ballistic relaxation at the beginning,
and go into a plateau regime, $\beta$-relaxation regime,
through a small dip, boson peak.
At the end, it shows a slow relaxation, $\alpha$-relaxation regime,
which is well described by a stretched exponential form [Eq.~(\ref{eq:str_exp1})].
Curves in Figs.~\ref{figs:sg}(a) and (b) show fitting results
for $\alpha$-relaxation.
The boson peak disappears when $\sigma \le 0.8$, and the autocorrelation
decays right after the ballistic relaxation.
Though the microscopic origin of the boson peak has been a debating issue,
it relates to relaxations in locally restricted area.
At $\sigma = 0.6$, the mean-field region,
the correlation hardly decays at the beginning,
and suddenly start decaying around $t \sim 1000$.
While the autocorrelation function cannot be fitted by exponential and power
functions, the Havriliak-Negami type~\cite{HavriliakNegami1967}
fitting function,
\begin{equation}
  C(t) = \frac{C_0}{[1 + (t/\tau)^\gamma]^x},
  \label{eq:havriliak_negami}
\end{equation}
well fits the data.
The resulting parameters are $\tau = 4.58\times 10^3$,
$\gamma = 1.69$, and $x = 4.24$, respectively.
The autocorrelation functions of the spin glass model
are different from those suggested by the droplet theory [see Sec.~\ref{sec:review_sg}].
Reasons are considered in the next section.

\section{%
  \label{sec:discussion}
  DISCUSSION
}

In this section, we first examine reasons of discrepancies
between the droplet theory and numerical data of pure
ferromagnetic model shown in Sec.~\ref{sec:result_pure}.
In Sec.~\ref{sec:result_dilute},
autocorrelation functions of the diluted ferromagnetic model
exhibit KT-like features, the power-law decay and a power-law
form of the structure factor.
Considering the origin of KT-like features,
though the ordered phase of the system seems KT phase,
it is revealed that the origin of the KT-like features is
a crossover effect caused by randomly diluted lattice.
The discrepancy between droplet theory
and numerical data in the spin glass model is discussed
in the last part of this section.

First we consider the reason why
two exponential functions are required
when we attempted to figure out what the function form
describes well numerical data of the pure ferromagnetic model
in $\sigma \le 1.6$.
The function form in the region
does not conform a result from the droplet theory,
that the autocorrelation function is described by
the simple exponential function when $d \ge 3$.
This threshold dimension is lower than the upper critical dimension, $d_u = 4$,
and the corresponding value of $\sigma$, therefore,
would be slightly larger than $\sigma_u = 3/2$.
The value of $\sigma$ at the threshold, $\sigma_c$,
is presumably close to 1.6,
and it is probable that
the change of the function form
relates to the intrinsic change of the function form
at the threshold.

\begin{figure}[h]
  \includegraphics[width=.4\textwidth]{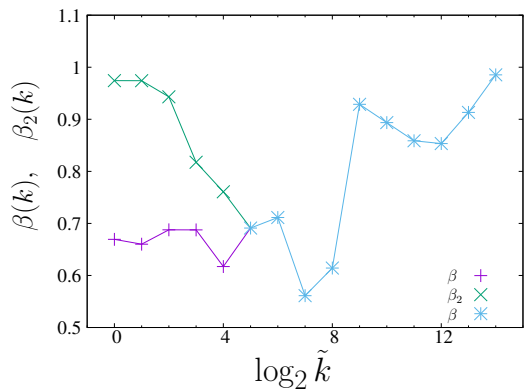}
  \caption{\label{fig:ferro_b1b2}
    (Color online) Plot of stretching exponents $\beta$ and $\beta_2$
    at $\sigma = 1.6$.
    While two stretching exponential functions are needed to fit data
    in $\tilde{k} \le 4$ [Eq.~(\ref{eq:str_exp2})],
    preparing a stretching exponential function is sufficient to fit data
    in $\tilde{k} \ge 5$.
  }
\end{figure}

The function form is the simple exponential function
in the mean-field model because the model reduces
a many-body dynamics to a one-body dynamics.
Therefore, it is naively expected that
autocorrelation functions are simple exponential
when $\sigma \le \sigma_u$.
However, autocorrelation functions at $\sigma=1.4$ and 1.5 (not shown)
are essentially the same as those at $\sigma=1.6$.
Figure~\ref{fig:ferro_b1b2} shows $\beta(\tilde{k})$ and
$\beta_2(\tilde{k})$ at $\sigma = 1.6$.
$\beta_2$ is nearly unity when $\tilde{k}$ is small,
it decreases as increasing $\tilde{k}$,
and it merges to $\beta$ at $\tilde{k} = 5$.
The stretched exponential function with $\beta$
seems an extra relaxation function
if we assume the stretched exponential function with
$\beta_2$ is derived from the simple exponential function
predicted by the droplet theory.
The extra unexpected relaxation would come from
the ununiformity in interactions which form surface of droplet.
While the droplet theory supposes the surface dimension of droplets
is $d-1$,
the surface dimension of droplets in our 1D model
is not obvious;
a plausible definition is a value of integral of interactions
which extend outward from a droplet,
\begin{equation}
  I = \int_{i \in \mathcal{C} \wedge j \notin \mathcal{C}} \frac{J}{r_{ij}^\sigma}dr_{ij},
\end{equation}
where $i$ runs from one end of the droplet to the other end,
and $j$ runs all over the sites except on the droplet.
If the droplet consists of a sequentially aligned Ising spins,
the integral value is proportional to $c^{2-\sigma}$,
where $c$ is a volume of the droplet.
Assuming the volume is proportional to the power of $d$,
the surface dimension of the droplet is estimated as $d(2-\sigma)$.
The result seems reasonable at both extremities of $\sigma$;
the integral value is equal to a constant at $\sigma = 2$
and is proportional to $d$ at $\sigma = 1$.
However, the value is unexpectedly small
at the marginal boundary of the mean-field region, $\sigma = 3/2$.
At the boundary the effective dimension of the volume of droplets is four,
so that we obtain $d(2-\sigma) = 2$ as the surface dimension.
If the surface dimension is strictly $d(2-\sigma)$,
the dynamics should change at $\sigma = 3/2$;
the system exhibits the stretched exponential decay in $\sigma > 3/2$,
the simple exponential decay in $\sigma < 3/2$, respectively.
However, numerical data are inconsistent with the result.
The inconsistency is probably caused by the fluctuation
of the surface dimension of droplets.
The surface dimension, $d(2-\sigma)$, is estimated by
assuming that droplets consist of sequentially aligned Ising spins.
Actually most of droplets consist of
a lot of fractions of sequential Ising spin sites,
so that the surface dimension of these droplets are larger than
the estimated value.
Additionally, relatively small fractions of Ising spin sites
contribute to lowering the effective surface dimension
because their surfaces are softer and fluctuate widely.
As a result, effective surface dimension of droplets
has a distribution even though the value of $\sigma$ is fixed.
The distribution brings about two relaxation forms,
the simple and the stretched exponential forms
near the boundary ($\sigma = 3/2$).

Next we consider reasons that the disagreement in the stretching exponent
$\beta$ between numerical data and the droplet theory at $\sigma \lesssim 2$.
While the droplet theory gives
$\beta = (d-1)/2\,(\gtrsim 0$ at $\sigma \lesssim 2)$,
numerically obtained $\beta$'s are fairly larger than the expected value.
An apparent disagreement in numerical data
is that the relaxation at the beginning is considerably slow.
According to the droplet theory,
the logarithmic derivative of the function is
\begin{equation}
  \frac{d}{dt}\ln C(k; t) \propto -\beta\left(\frac{t}{\tau}\right)^{\beta-1}.
\end{equation}
Therefore, the slope of the autocorrelation function should be
steep at the beginning ($t \ll \tau$)
if $\beta$ is nearly zero as given by the droplet theory.
The absence of the steep decay is caused by the discreteness of
the Ising variable.
As shown in the inset of Fig.~\ref{figs:ferro}(a),
the emergence of small sized droplets are suppressed,
though the phase is supposed to be in the KT phase at $\sigma = 2$.
While the continuous spin variable as in the XY model
is able to contribute to the steep decay by
creation and annihilation of spin waves,
these small energy contributions are absent in the Ising spin system.
The effect of the discreteness is reduced when the temperature
is close enough to the critical point;
that is, thermal fluctuations blur the discreteness of Ising spin,
so that the structure factor is expressed by the power of the wave length.
In such a nearly critical system, the steep decay,
which coincides with a power-law decay at the criticality,
should be observed
since large $k$ droplet excitations decay at the beginning.
Since the system at $\sigma 1.9$ has the true long-range order,
the tail of the structure factor is not long as that of the KT phase.
However, a small stretching exponent would be observed
if the temperature of the ordered phase is high enough to
blur the discreteness of the Ising spin.

The autocorrelation functions of the diluted model
show the power-law decay and the fractal-like structure
though the system is in the ferromagnetic ordered state.
It is known that
these pseudo critical features are caused by
randomly diluted lattice~\cite{Ikeda1993}.
When the degree of dilution is slightly lower than the critical point,
the correlation length between lattice sites is given by
\begin{equation}
  \xi_p \sim (p - p_c)^{-\nu_p},
\end{equation}
where $p$ is the concentration of lattice sites,
$p_c$ is the critical concentration of lattice sites,
and $\nu_p$ is the critical exponent of the percolation correlation length.
The ordered region is proportional to the number of lattice sites,
when ferromagnetic regions are larger than $\xi_p$.
On the other hand, the ordered region forms a fractal structure,
when ferromagnetic regions are smaller than $\xi_p$.
Thus, while the structure factor shows the Lorentzian peak shape
in $k < 1/\xi_p$,
it shows the power-law shape in $k > 1/\xi_p$.
The structure factor of the diluted model seems to be classified
into three regions.
As an example, the structure factor at $\sigma = 1.6$
in the inset of Fig.~\ref{figs:dilute}(c)
is classified as follows:
(i) it is suppressed by the discreteness of Ising spin in $\tilde{k} \ge 12$,
(ii) it is almost a constant in $\tilde{k} \le 2$, and
(iii) it shows the power-law shape in $2 < \tilde{k} < 11$.
This classification indicates
the percolation correlation length is estimated as roughly $L/8$
when $\sigma = 1.6$.
The relaxation also depends on $k$;
relaxations of autocorrelations in $\tilde{k} \le 2$ are considerably
slower than those in $\tilde{k} > 2$ [see Fig.~\ref{figs:dilute}(c)
and Fig.~\ref{figs:dilute_ttx}(c)].
Essentially the same behaviors are also observed at $\sigma = 1.4$.
Though effects of the random dilution is averaged out in
the mean-field limit, as discussed above,
fractal-like features still remain at $\sigma = 1.4$
due to the distribution of the surface dimension of droplet.
Whereas slow decay in autocorrelation functions of $\tilde{k} < 1/\xi_p$
is observed at $\sigma \le 1.6$,
there is no such a signal at $\sigma > 1.6$.
The difference is caused by that $\xi_p$ becomes larger as increasing $\sigma$.
The transition temperature decreases as increasing $\sigma$,
and thermal fluctuations become weaker.
The system comes close to the percolation transition
point as decreasing the transition temperature,
and the approaching to the point causes the growth of $\xi_p$.
As a result, the autocorrelation functions at $\sigma > 1.6$
seems like fractal in all the region of $\tilde{k}$.

The fractal-like features are also observed in
disordered ferroelectrics.
Koreeda and collaborators observed the power-law distribution
in the quasielastic light scattering in Pb(Mn$_{1/3}$Nb$_{2/3}$)O$_3$,
and they also observed the power exponent of the spectrum depends
on temperature~\cite{Koreeda2012}.
Their observations are indeed the same as those of results that
the droplet theory gives.
Dynamics of disordered ferroelectrics are not understood well.
To study of the dynamics, spin glass models are often employed
since they exhibit spin glass like behaviors,
extremely slow dynamics which involve aging, memory effect,
frequency dependence of the AC susceptibility, \textit{et al}.
However, considering the experimental results by Koreeda and collaborators
and the results from diluted ferromagnetic models,
it should be examined which models is proper to describe phenomena of interest.

Autocorrelation functions of the spin glass model
obtained by the numerical simulation
are quite different from those given by the droplet theory.
While their decay form is given by the power of logarithmic function of time
according to the droplet theory,
numerical data exhibit the stretched exponential decay in the range
of $0.7 \le \sigma \le 1$ and Havriliak-Negami type 
relaxation at $\sigma = 0.6$ [Eq.~(\ref{eq:havriliak_negami})].
The disagreement would come from an intrinsic difference between
microscopic and macroscopic variables, which makes little difference
in ferromagnetic phase.
While, in this paper,
the Edwards-Anderson (EA) order parameter~\cite{EdwardsAnderson1975}
[Eq.~(\ref{eq:edwards_anderson})]
is employed as a macroscopic variable to investigate dynamical properties,
the droplet theory gives the time-delayed correlation of a local spin variable.
The EA order parameter measures a similarity between
replicated systems which have the same set of interactions $\{J_{ij}\}$.
Therefore, the autocorrelation of the EA order parameter
measures the time-delayed correlation of the similarity.
On the other hand, the time-delayed correlation of a local spin variable
measures a local spin dynamics in a sample.
Thus, autocorrelations of the two observables exhibit different
relaxations.

\section{%
  \label{sec:summary}
  SUMMARY
}

In this paper, Monte Carlo simulations on 1D Ising models
with long-range interactions are executed,
and the numerical data are compared with results given by the droplet theory.
Essentially consistent results are obtained in the pure
and the diluted ferromagnetic models,
though disagreements caused by the discreteness of Ising spin
and the distribution of the surface dimensionality
are also observed.
Numerical results in the spin glass model are different from
results given by the droplet theory.
This means that dynamics of the macroscopic variable
are essentially different from
those of the microscopic variable analyzed in the droplet theory.

It is shown that
combined use of the droplet theory and numerical simulation
helps us to understand nature of ordered states.
Applying the results to investigations of non-trivial phases
as in disordered ferroelectrics~\cite{Koreeda2012}
or frustrated magnets~\cite{Shinaoka2014}
could advance understanding of ordered states.
To extend the range of applicable fields,
clarifying relations between ordered states and its dynamics
in other fundamental spin models is required.

\begin{acknowledgments}
  The author thanks Prof. Takayama and Dr. Nonomura for useful comments
  on the manuscript.
  The random-number generator MTGP23209~\cite{SaitoMatsumoto2013}
  was used for numerical calculations.
\end{acknowledgments}

\appendix*
\section{%
  \label{sec:appendix}
  $O(N)$ Monte Carlo method
}

The main idea of the $O(N)$ MC method is to estimate
a molecular field at site $i$ by sampling interacting sites.
The Hamiltonian of the Ising model with long-range interactions is
given by
\begin{equation}
  \label{eq:hamiltonian_apx}
  {\cal H} = -\sum_{i < j}J_{ij}S_iS_j.
\end{equation}
The summation in Eq.~(\ref{eq:hamiltonian_apx}) runs over every pair
of interactions: therefore, interactions $J_{ij}$ should not be
those of bare but of rescaled
when the periodic boundary condition is imposed.
We suppose all the interactions are ferromagnetic ($J_{ij} \ge 0$)
for convenience in explaining.
The extension to spin glass model is given at the bottom of Appendix.

The Boltzmann weight $W_{\rm B}(\{S\})$ for site $i$ is given by
\begin{align}
  W_{\rm B}(\{S\}) &= \prod_{j(\ne i)}\exp(\beta J_{ij}S_iS_j)\nonumber\\
  &= \prod_{j(\ne i)}\left[
    \frac{1+S_iS_j}{2}e^{\beta J_{ij}} + \frac{1-S_iS_j}{2}e^{-\beta J_{ij}}
    \right],
\end{align}
where $\{S\}$ represents a spin configuration.
By introducing an auxiliary parameter $\alpha(> 0)$,
we are able to deactivate interactions stochastically,
and the deactivation reduces the computational cost
with the stochastic legitimacy.
After introducing the parameter $\alpha$,
the resulting $W_{\rm B}(\{S\})$ is written by~\cite{Tomita2009a}
\begin{equation}
  \label{eq:onbw}
  W_{\rm B}(\{S\}) \propto \sum_{\{k\}}
  P(\lambda_{\rm tot}; k_{\rm tot})
  f(\{k\}; \{\lambda\})
  w(\{k\}; \{\lambda\}; \{S\}).
\end{equation}
Here,
$P(\lambda_{\rm tot}; k_{\rm tot})$,
$f(\{k\}; \{\lambda\})$, and
$w(\{k\}; \{\lambda\}; \{S\})$
are the Poisson probability mass function,
the multinomial probability mass function,
and a weight function, respectively.
The explicit forms of functions are as follows:
\begin{widetext}
\begin{align}
  \label{eq:poisson}
  & P(\lambda_{\textrm{tot}}; k_{\textrm{tot}}) =
  e^{-\lambda_{\textrm{tot}}}\frac{\lambda_{\textrm{tot}}^{k_{\textrm{tot}}}}{k_{\textrm{tot}}!},\\
  & f(\{k\}; \{\lambda\}) =
  k_{\rm tot}!
  \prod_{j(\ne i)}\left[\frac{1}{k_{ij}!}
    \left(\frac{\lambda_{ij}}{\lambda_{\rm tot}}\right)^{k_{ij}}
    \right],\\
  & w(\{k\}; \{\lambda\}; \{S\}) =
  \prod_{j(\ne i)}\left[
    \frac{1 + S_iS_j}{2} + \frac{1 - S_iS_j}{2}
    \left(\frac{\alpha}{\lambda_{ij}}\right)^{k_{ij}}
    \right],
\end{align}
\end{widetext}
where $\lambda_{ij} = 2\beta J_{ij} + \alpha$,
$\lambda_{\rm tot} = \sum_{j(\ne i)}\lambda_{ij}$,
and $k_{\rm tot} = \sum_{j(\ne i)}k_{ij}$.
The variable $k_{ij}$ is a number of activated bonds
between $i$ and $j$,
and a set $\{k\}$ represents an activated bond configuration.

$P(\lambda_{\rm tot}; k_{\rm tot})$ gives a stochastic weight
of a number of total bonds, $k_{\rm tot}$,
whose mean value is $\lambda_{\rm tot}$.
The constant $\lambda_{\rm tot}$ is the summation of
an effective interaction $\lambda_{ij}$.
In the $O(N)$ MC method, an effective interaction $\lambda_{ij}$ appears
as a summation of the bare interaction $\beta J_{ij}$
and an auxiliary tunable interaction $\alpha$.
The auxiliary parameter $\alpha$ determines the efficiency of
the $O(N)$ MC method.
Decreasing the value of $\alpha$ decreases the number of active bonds,
and it reduces the computational cost relating to the bond activation.
However, small $\alpha$ brings about a low acceptance ratio for
a spin flip.
Indeed, in the limit of $\alpha = 0$,
the $O(N)$ MC method corresponds to the Swendsen-Wang cluster
MC method~\cite{SwendsenWang1987}.
Conversely, increasing the value of $\alpha$ raises an acceptance ratio
for a spin flip, but it also raises the number of active bonds
and the computational cost.
In fact, the limit of $\alpha = \infty$ reduces the $O(N)$ method
to the Metropolis method.
The dynamics of the simulation is optimized by tuning the parameter $\alpha$.
A reasonable choice of $\alpha$ is
\begin{equation}
  \label{eq:alpha}
  \alpha_{ij} = 2\beta J_{ij}\tilde{\alpha},
\end{equation}
where $\tilde{\alpha}$ is a constant.
This choice makes $\lambda_{ij}$ proportional to $\beta J_{ij}$ as
\begin{equation}
  \label{eq:lambda}
  \lambda_{ij} = 2\beta J_{ij}(1 + \tilde{\alpha});
\end{equation}
this means that all of the bonds are activated with probabilities
proportional to the bare interaction $\beta J_{ij}$.
Using the choice, functions $f(\{k\}; \{\lambda\})$ and
$w(\{k\}; \{\lambda\}; \{S\})$ are rewritten, respectively, by
\begin{align}
  \label{eq:multinomial}
  f(\{k\}; \{\lambda\}) &=
  k_{\rm tot}!
  \prod_{j(\ne i)}\left[\frac{1}{k_{ij}!}
    \left(\frac{J_{ij}}{J_{\rm tot}}\right)^{k_{ij}}
    \right],\\
  w(\{k\}; \{\lambda\}; \{S\}) &=
  \prod_{j(\ne i)}\left[
    \frac{1 + S_iS_j}{2} + \frac{1 - S_iS_j}{2}\kappa^{k_{ij}}
    \right]\nonumber \\
  \label{eq:wstochastic}
  &= \prod_{j\in (S_j\nshortparallel S_i)}\kappa^{k_{ij}}
\end{align}
where $J_{\rm tot} = \sum_{j(\ne i)}J_{ij}$ and
$\kappa = \tilde{\alpha}/(1 + \tilde{\alpha})$,
and the product
in Eq.~(\ref{eq:wstochastic}) runs over antiparallel spin pairs.
Equation~(\ref{eq:multinomial}) gives a stochastic distribution
so that $k_{ij} \propto J_{ij}$.
A stochastic weight of a spin configuration $\{S\}$
in a bond configuration $\{k\}$ is given by Eq.~(\ref{eq:wstochastic}).

Using the Eqs.~(\ref{eq:poisson}), (\ref{eq:multinomial}), and (\ref{eq:wstochastic}),
a pseudocode of a heat-bath spin update with the $O(N)$ MC method
is implemented as follows:
\begin{quote}
  $p\coloneqq 1$\\
  $q\coloneqq 1$\\
  $c_p^{\shortparallel}\coloneqq 2\tilde{\alpha}+2$\\
  $c_q^{\shortparallel}\coloneqq 2\tilde{\alpha}$\\
  $c_p^{\nshortparallel}\coloneqq 2\tilde{\alpha}^2+2\tilde{\alpha}+1$\\
  $c_q^{\nshortparallel}\coloneqq 2\tilde{\alpha}^2+4\tilde{\alpha}+1$\\
  $k_{\textrm{tot}}\coloneqq $Poisson($\lambda_{\textrm{tot}}$)\\
  \textbf{for} $k=1$ \textbf{to} $k_{\textrm{tot}}$ \textbf{do}\\
  \hspace{2em}$j\coloneqq$ Walker($\{\lambda\}$)\\
  \hspace{2em}\textbf{if} $S_j = S_i$ \textbf{then}\\
  \hspace{4em}$p\coloneqq c_p^{\shortparallel}p$\\
  \hspace{4em}$q\coloneqq c_q^{\shortparallel}q$\\
  \hspace{2em}\textbf{else}\\
  \hspace{4em}$p\coloneqq c_p^{\nshortparallel}p$\\
  \hspace{4em}$q\coloneqq c_q^{\nshortparallel}q$\\
  \hspace{2em}\textbf{end if}\\
  \textbf{end for}\\
  \textbf{if} Random() $> \frac{p}{p+q}$ \textbf{then} $S_i \coloneqq -S_i$
\end{quote}
A formula $a \coloneqq n$ means that $n$ is plugged in for $a$.
Poisson($\lambda_{\textrm{tot}}$) generates a random number from a Poisson
distribution with a mean of $\lambda_{\textrm{tot}}$,
Walker($\{\lambda\}$) generates a random number from
a distribution given by Eq.~(\ref{eq:multinomial})~\cite{Walker1977}.
Random() generates a random number from a uniform distribution
in the range from 0 to 1.

Since the cost of $O(N)$ MC method for interaction pairs
which give $\lambda_{ij} > 1$ is more expensive than
that of the conventional MC method,
switching the stochastic weight from the one of the $O(N)$ MC method
to the conventional one reduces the cost of computation.
For the combination use, we introduce an arbitrary constant $c$
and separate interacting pairs into $\{ij\}_{< c}$,
a set of pairs giving $\lambda_{ij} < c$,
and $\{ij\}_{\ge c}$,
the set of remaining pairs giving $\lambda_{ij} \ge c$.
Then, 
the spin flip probability, $p/(p+q)$,
is replaced by $p/(p+q\exp\{-\beta\Delta E\})$,
where $\Delta E$ is the single-spin-flip energy difference
for $\{ij\}_{\ge c}$.

When an interaction $J_{ij}$ is randomly diluted to
$\gamma J_{ij} (0 \le \gamma < 1)$,
the multinomial probability mass function, $f(\{k\}; \{\lambda\})$,
becomes dependent on the site $i$.
If we modify the function for each site $i$,
a large memory area which is proportional to $O(N^2)$ is needed.
We can save the large memory area by 
changing the multiplicative factors, $c$'s in the pseudocode, as
\begin{align*}
  c_p^{\shortparallel} &\coloneqq 2\tilde{\alpha}+1+\gamma,\\
  c_q^{\shortparallel} &\coloneqq 2\tilde{\alpha}+1-\gamma,\\
  c_p^{\nshortparallel} &\coloneqq 2\tilde{\alpha}^2+(3-\gamma)\tilde{\alpha}+1,\\
  c_q^{\nshortparallel} &\coloneqq 2\tilde{\alpha}^2+(3+\gamma)\tilde{\alpha}+1.
\end{align*}
The constant $\gamma$ substitutes the change in $f(\{k\}; \{\lambda\})$
with the spin-flip probability.
Employing the modification of the multiplicative factors,
we can use the same $f(\{k\}; \{\lambda\})$ as the pure system
for the diluted spin system.

For the antiferromagnetic system, the satisfied spin configuration
inverts from parallel to antiparallel,
so that the weight function for the system is
converted to
\begin{equation}
  \label{eq:wstochastic2}
  w(\{k\}; \{\lambda\}; \{S\}) =
  \prod_{j(\ne i)}\left[
    \frac{1 + S_iS_j}{2}\kappa^{k_{ij}} + \frac{1 - S_iS_j}{2}
    \right].
\end{equation}
For the spin glass model, we apply Eq.~(\ref{eq:wstochastic})
and Eq.~(\ref{eq:wstochastic2}) for ferromagnetic
and antiferromagnetic interactions, respectively.
    
\bibliography{tomita}

\end{document}